\documentclass[final,5p,times,twocolumn]{elsarticle}

\usepackage{graphics,graphicx,dcolumn,bm,fleqn,epic,eepic,float,ulem}
\usepackage{amssymb,amsmath,multirow,rotate,color}

\definecolor{red}{rgb}{1,0,0}
\definecolor{green}{rgb}{0,1,0}
\definecolor{blue}{rgb}{0,0,1}

\newcommand{\ca}{Ca$^{2+}$}

\journal{Journal of Theoretical Biology}
\begin{document}

\begin{frontmatter}

\title{Modeling the functional network of  primary intercellular Ca$^{2+}$ wave propagation in  astrocytes and its application to study drug effects}

\author{Marcelo Pires$^{a,b}$}
\author{Frank Raischel$^{a,c}$}
\author{Sandra H.~Vaz$^{d,e}$}
\author{Andreia Cruz-Silva$^{d,e}$}
\author{Ana M.~Sebasti\~ao$^{d,e}$}
\author{Pedro G.~Lind$^{a,f}$}

\address{$^a$Centro de F\'{\i}sica Te\'orica e Computacional,
             Faculdade de Ci\^encias,
             Universidade de Lisboa,
             Campo Grande 1649-003 Lisboa, Portugal}
\address{$^b$Departamento de F\'{\i}sica, 
             Universidade Federal do Amap\'a, 
             Jardim Marco Zero, 68903-419 Macap\'a/AP, Brazil}
\address{$^c$Centro de Geof\'{\i}sica, Instituto Dom Luiz,
         Universidade de Lisboa,
         1749-016 Lisboa, Portugal}
\address{$^d$Instituto de Farmacologia e Neuroci\^encias,
             Faculdade de Medicina,
             Universidade de Lisboa,
             1649-028 Lisboa, Portugal}
\address{$^e$Unidade de Neuroci\^encias, 
             Instituto de Medicina Molecular,
             Universidade de Lisboa,
             1649-028 Lisboa, Portugal}
\address{$^f$ Institute f\" ur Physik and {\it ForWind},
              Carl-von-Ossietzky Universit\"at Oldenburg, DE-26111 Oldenburg, 
              Germany}

\begin{abstract}
We introduce a simple procedure of multivariate signal analysis to uncover 
the functional connectivity 
among cells composing a living tissue and describe 
how to apply it for extracting insight on the effect of  drugs in the tissue.
The procedure is based on the covariance matrix of time resolved activity 
signals.
By determining the time-lag that maximizes covariance, one derives the weight 
of the corresponding connection between cells. 
Introducing simple constraints, it is possible to conclude whether pairs of 
cells are functionally connected and in which direction. 
After testing the method against synthetic data we apply it to study 
intercellular propagation of Ca$^{2+}$ waves in astrocytes 
following an external stimulus, with the aim of 
uncovering the functional cellular connectivity network.
Our method proves to be particularly suited for this type of networking 
signal propagation where signals are pulse-like and have short time-delays, 
and is shown to be superior to standard methods, namely a multivariate 
Granger algorithm.
Finally, based the statistical analysis of the connection weight 
distribution, we propose simple measures for assessing the impact of drugs
on the functional connectivity between cells.
\end{abstract}


\end{frontmatter}


\section{Introduction}
\label{sec:introducao}

Astrocytes, which were long thought to perform only auxiliary 
functions in the brain, are known to exhibit complex patterns 
of Ca$^{2+}$ waves propagating in their cellular network\cite{newman2001}.
The basic biological mechanisms that underlies the functional links between 
astrocytes, leading to consecutive elevations in calcium 
signal\cite{haydon2001,Devinsky2013}.
The predominant mechanism is mediated by ATP, which activates metabotropic 
P2Y receptors in the astrocytic membrane, leading to the formation of 
inositol-3-phosphate (IP$_3$), which then signals to release calcium from the 
intracellular stores.
Calcium elevation in an astrocyte leads to further release of ATP to the 
extracellular media, which quickly acts in receptors in the membrane of 
neighboring cells, leading to calcium elevations in those cells, which 
leads to a continuous cascade of \ca signal propagation.
Transfer of IP$_3$ across gap-junctions (connexins) may also contribute for
the calcium elevation\cite{hoefer2002}, though in a minor 
degree\cite{haydon2001,hassinger1996,bennett2005}.
Other recent studies have shown that intracellular Ca$^{2+}$ oscillations
are  basically a form of correlated noise\cite{perc2008}, 
which raises the question of stochasticity and reproducibility of the 
signal propagation and the corresponding network.
While questions concerning the details of the propagating mechanisms
remain an important matter of discussion\cite{falcke2004}, we are here
interested in extracting insight from the inter-connectivity between
cells by keeping track of Ca$^{2+}$ signals.
\begin{figure}[t]
\centering
\includegraphics[width=0.43\textwidth]{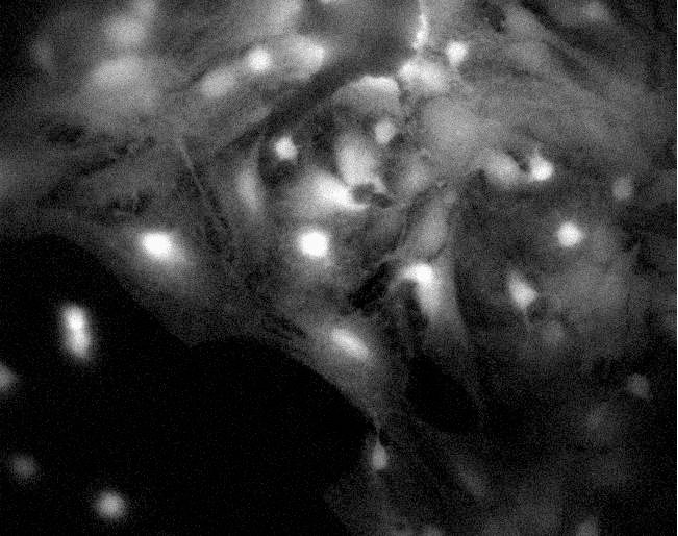}
\includegraphics[width=0.43\textwidth]{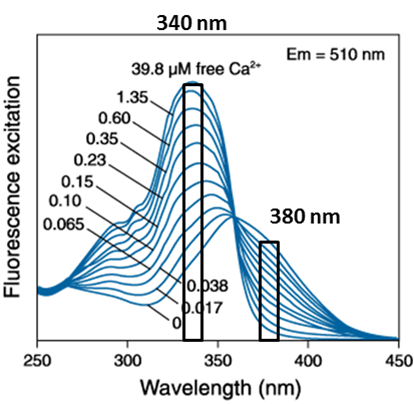}
\caption{\protect 
         Set of $20$ astrocytes observed through calcium imaging procedures.
         Bright regions indicate the location of cytoplasm and organelles, 
         where the 
         concentration of Ca$^{2+}$ is
         higher than in the dark regions indicating the intercellular
         medium, where diffusion processes take place. 
         Concentration is measured by the ratio $R$ between the 
         radiation emitted at $510$ nm when cells are excited at $340$ nm 
         over emission upon excitation at $380$ nm.}
\label{fig1}
\end{figure} 

Uncovering the functional connectivity of astrocytes in these networks
provides a better understanding of the functionality
of the astrocytic network itself, but also allows us to assess
drug effects, not only at the single cell level, but also upon
the spreading of the signal throughout the cellular 
tissue\cite{falcke2004}.
Such activity can be well characterized by measuring the concentration of 
calcium ions (Ca$^{2+}$) using noninvasive techniques of calcium 
imaging.
One of the most popular of such techniques uses fluorescent dye 
indicators, which bind selectively to free Ca$^{2+}$ ions, undergoing a 
conformational change and consequently a variation in its fluorescence 
excitation and/or emission properties when bounded to Ca$^{2+}$. 
These variations can be used to evaluate changes in intracellular 
Ca$^{2+}$ concentration.
Here concentration is measured by the ratio $R$ between maximum amplitudes
at $340$ and $380$ nm, as illustrated in Fig.~\ref{fig1}.

The evolution of Ca$^{2+}$ concentration in each cell depends typically 
on the diffusion of the signaling molecules through the intercellular 
environment and on the direct connection from one astrocyte to its neighbors.
While the former mechanism is slow, the latter is fast and dominates
for measurement series with high sample rates. This study focuses on the
latter case.
Therefore, the series of measurements of Ca$^{2+}$ concentrations at each 
cellular location reflects the flow of Ca$^{2+}$, and consequently the 
propagation of this ion from each cell to the neighboring ones.
It should be noted that the brightening front generated at each cell does
not necessarily propagate radially to its neighbors, since the physical
connections between neighboring cells are heterogeneously distributed, 
which introduces spatial  inhomogeneities  in the signal
propagation\cite{hoefer2002,giaume1998}.

\begin{figure*}[t]
\centering
\includegraphics[width=0.95\textwidth]{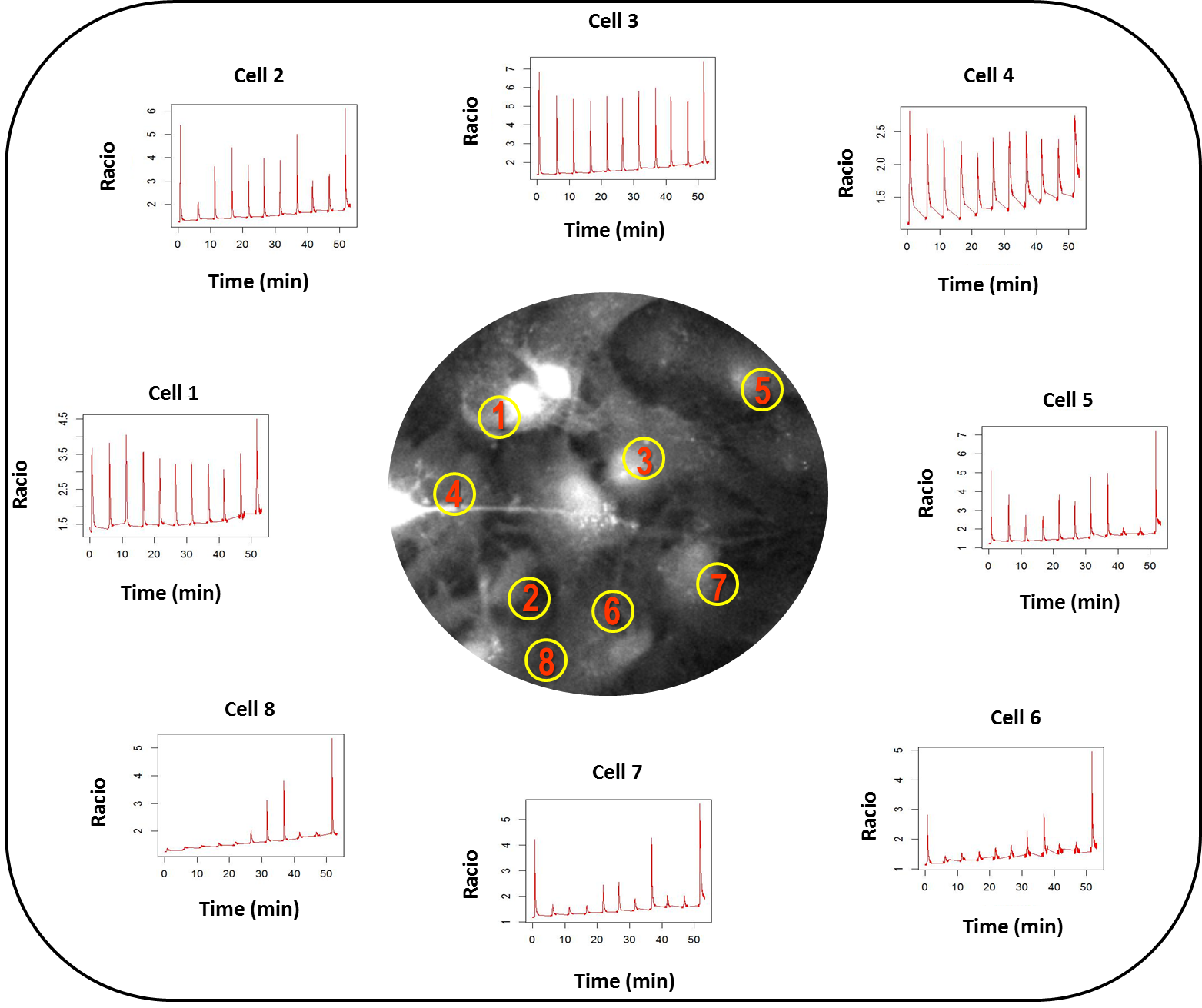}
\caption{\protect 
         Intercellular \ca signaling cascade.
         When one has a set of time series of the radiation ratio $R$ 
         measured at each cell of a tissue sample, how can one 
         infer the flow  of Ca$^{2+}$ through the tissue? 
         The central plot shows a case of eight signals extracted from the
         indicated positions, one per cell, forming the culture of confluent
         astrocytes that is used as a tissue model.
         In the ten surrounding signal plots, each plot shows
         the respective response of an individual cell to ten different 
         ATP stimuli (see text) applied to  cell  $1$.
         The final, eleventh peak shows  the maximum cellular response,
         recorded by applying a supramaximal ATP stimulus (100uM) to cell $1$,
         and was not used in the analysis below.}
\label{fig2}
\end{figure*}

One possible way to model \ca signal propagation is through a bottom-up 
approach, where the kinetic constants of intracellular and intercellular 
signal pathways, the spatial  distributions of connections and receptors 
in the tissue, and the relative importance of signal propagation mechanisms   
needs to be known and incorporated in a detailed mechanical simulation. 
However, this approach is quite cumbersome and case dependent.

We argue, however, that the detailed physical structure of interconnected 
cells is not necessary to characterize the response of the interconnected 
tissue to  a signal.
For that one only needs to uncover the so-called functional connectivity
between cells, which describes the synchronization patterns between
the cells.
While functional connectivity is at most loosely related to causal 
connectivity, it reflects the way the tissue  as an
interconnected structure of cells, responds to external stimuli, 
and how this functional network is changed when applying  drugs.

Therefore, for our purposes, we focus here on the functional connectivity.
Similar approaches to assess the functional connectivity in biological
systems have already been presented, e.g.~in islets of Langerhans from
mouse pancreas tissue slices\cite{PLoS2013}.
 
From the mathematical perspective, we quantify the connectivity and causality 
of information flow between the cells as a weighted graph.
A graph is a collection of nodes  interconnecting through edges according to 
some specific  rules \cite{Diestel2005}.  

Our aim here is  mapping the complex interactions and temporal information 
flow patterns to a much simpler representation.
The graph is completely characterized by a single matrix, 
the adjacency matrix, whose entries are the weights of the edges between nodes.
In case of cellular tissues, nodes represent the cells, and edges between 
them the functional connections between a respective pair of cells.

To reconstruct this graph we consider the paths connecting neighboring 
astrocytes in a confluent culture of astrocytes, used as a  tissue model, 
analyzing the time series of 
Ca$^{2+}$ concentration measurements observed at each astrocyte separately
(see Fig.~\ref{fig2}). From these experimentally measured 
synchronization patterns, we employ the correlation measures between the \ca
time series.

Standard algorithms exist for this problem , based in Granger 
causality\cite{granger1969,blino2004,dhamala2008}.
However,  although they can be adapted to suit multivariate non-stationary data 
series\cite{granger1964,granger1986}, these methodologies assume the existence of 
a random process with a certain level of stationarity and then apply spectral 
analysis, decomposing the data series into a certain number of uncorrelated 
components, each one corresponding to a given spectral frequency.

For our particular case where one has sets of externally stimulated signals
that are typically pulse-like and with short time-delays
between them, a different approach must be considered, as will become apparent 
below.

Here, we show that these particular signals  observed in signal propagation 
of astrocytic tissues can best be recovered by our  simple and accurate 
procedure, which  can also be applied for assessing the efficiency and 
irreversibility of  drug infusion on the structure of the tissue, by 
studying signal propagation.

This report is organized as follows. We start in Section \ref{sec:model} 
by describing the general properties of the signal and then give a detailed 
account of our procedure.  In particular, we argue that using the normalized 
covariance between astrocytes for specific time-lags enables one to quantify 
the connectivity of each pair of astrocytes. 
We then show that, for synthetic  networks, our procedure is more efficient 
and accurate than standard algorithms such as Granger causality
In Sec.~\ref{sec:data} we describe the experimental setup and the data extracted from samples of living astrocyte tissues. 
In Sec.~\ref{sec:analysis}, we extract the connectivity networks for each 
sample and for each stimulus, using our procedure, and afterward analyze the moments of 
the weight distribution found for each tissue sample to discuss the 
effect of two drugs on the connectivity structure and the signal 
propagation, and the variability of the results.
Finally, Sec.~\ref{sec:conclusions} concludes the paper giving also a
brief description of  how this method can be applied to drug tests.
Our auxiliary model for creating synthetic signal data is described in \ref{appendix:synth}.
The Granger method and algorithm is   briefly described in \ref{appendix:granger}.

\section{Extracting cellular functional connectivity}
\label{sec:model}

\subsection{Properties of the experimental signal propagation}
\label{subsec:exp_prop}

A brief description of the common numerical features of the \ca cascade follows. A 
detailed description of the experimental setup can be found in Sec.~\ref{sec:data}.

In all experiments,  cell 1 is externally stimulated  10 times,
with $10\ \mu$M ATP (focally applied for $200$ ms), the 11th peak 
being induced by a supramaximal concentration of ATP ($100\ \mu$M) to test 
the maximal activity of the cell.
The  \ca signal cascade is observed by measuring the intracellular  \ca levels, 
cf. Figs.~\ref{fig1},\ref{fig2}, through the radiation amplitude ratio $R$.

The measured temporal signals, cf.~Fig.~\ref{fig2}, are a series of pulses, accompanied 
by an increasing trend, the latter not being of interest here. 
The signals are of similar shape, which is  initially  Gaussian,  before decaying 
with a slower-than exponential  tail. The amplitudes, however, are widely varying. 
Generally speaking, it can be observed  in this example that the stimulated cell 1---and 
also cell 4---have more or less constant
signal amplitude ratios, whereas the cells farther away from cell 1 exhibit lower amplitudes
and less regular peak heights. The signals are all delayed with respect to cell 1, although
this crucial property is not visible at the resolution level provided by Fig.~\ref{fig1}.       
This delay, specifically the delay in correlations related to it, is the cornerstone of 
our method of reconstructing the signal network. 

\subsection{Modeling  functional connectivity strength from signal correlation}
\label{subsec:sig_corr}

\begin{figure}[htb]
\centering
\includegraphics[width=0.45\textwidth]{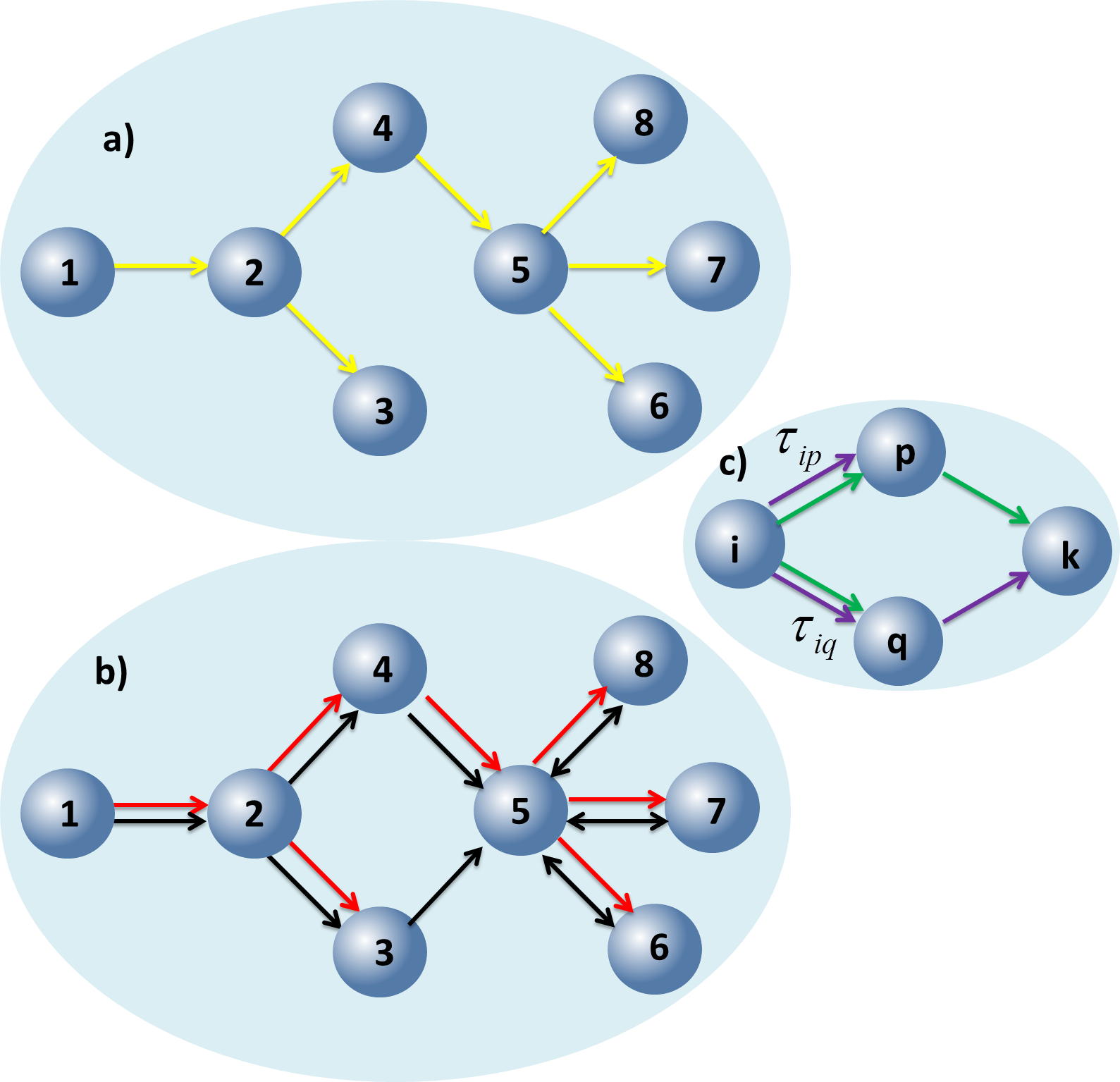}
\caption{\protect 
         {\bf (a)}
         A sample directed graph of functional connectivity  as it can 
         be extracted by our method, here for $M=8$ interconnected 
         (artificial) astrocytes.
         Nodes represent the cells (blue bullets) and edges represent 
         their  connections (yellow arrows). 
         {\bf (b)} The connections can be evaluated from the 
         time series at each node using two different procedures: 
         Granger causality\cite{granger1969,blino2004,dhamala2008} 
         (black arrows) or our method (red arrows).
         Comparing the results with the synthetic network of astrocytes
         in (a), our method can be taken as better for this particular
         purpose (see text and Fig.~\ref{fig7}).
         {\bf (c)} Particular cases of signal correlations, namely
         Pearson coefficient equal to one, may represent different but 
         equivalent signal propagation topologies.} 
\label{fig3}
\end{figure}

Following the experimental results, we propose a  model to obtain the network of
functional connectivity, i.~e.~the existence, direction and strength of network 
links between the network nodes, which are the cells, from the measured signal 
correlations. The result is then a network comparable to Fig.~\ref{fig3}a.

We consider a number $M$ of cells from which the Ca$^{2+}$ can be measured
composing the time series $X_i(t)$, $i=1,\dots,M$ and $t$ labeling
time-steps.
To derive the connectivity between a pair of cells, say $i$ and $j$, 
we consider primarily how  strong the corresponding signals, $X_i(t)$ 
and $X_j(t)$, are correlated. 
Since we are dealing with signal propagation, one must also consider a 
time-delay $\tau$ separating the two measures $X_i(t)$ and $X_j(t-\tau)$.
In particular, we assume that a proper choice of the  value of $\tau$ 
may completely reproduce at cell $j$  the shape of the signal occurring 
previously in cell $i$, i.e.~$X_i(t)=X_j(t-\tau)$ for all times 
$t=n\Delta t$, modulo an attenuation factor.
Here, $\Delta t$ is the inverse of the sample rate taken for
extracting the series of measurements composing the signal 
at steps $n=1,\dots,N$, with $N$ indicating the total number of measurements.

Typically, the signal at one cell is not completely reproduced in another
cell, since the  signal propagation involves various mechanisms, such as 
gap junction transport and diffusion.
To determine how strong the signals are correlated, we take their 
covariance with a time-delay $\tau$:
\begin{eqnarray}
C(X_{i}, X_{j},\tau) = \frac{  \sum_{t=\tau}^{N} 
                  ({X}_{i}(t)-\bar{X}_{i})({X}_{j}(t-\tau)-\bar{X}_{j})} 
                  {(N-1)\sigma_i\sigma_j} 
\label{eq:cov}
\end{eqnarray} 
where $\bar{X}_{i}$  and  $\sigma_i$ stand for the average and standard 
deviation of signal $X_i(t)$ respectively.
This measure has the properties
$C(X_i,X_i,0)=1$ and $\lim_{\tau\to \infty}C(X_i,X_j,\tau)=0$.

The value of the covariance between the signals at two neighboring
cells may vary depending on the type of stimulus applied at the 
source-cell. Also, when repeatedly applying the same stimulus to the same 
array of cells, signal propagation may be altered by aging or learning 
effects. One would also expect that the effect of drugs 
is reflected by the covariance, such that, when the covariance increases in 
absolute value, it could indicate an excitatory effect of the stimulus 
substance. Or, when it decreases,  it could reveal an inhibitory effect. 

We stress once more that, 
while it remains only a hypothesis that the strength of the connection
between two cells can be ascertained from the covariance
their functional connectivity is indeed reflected in
the covariance: 
the larger the covariance between two signals is in absolute value, the 
stronger the connection between the functional behavior of the 
corresponding cells should be.

\begin{figure}[htb]
\centering
\includegraphics[width=0.45\textwidth]{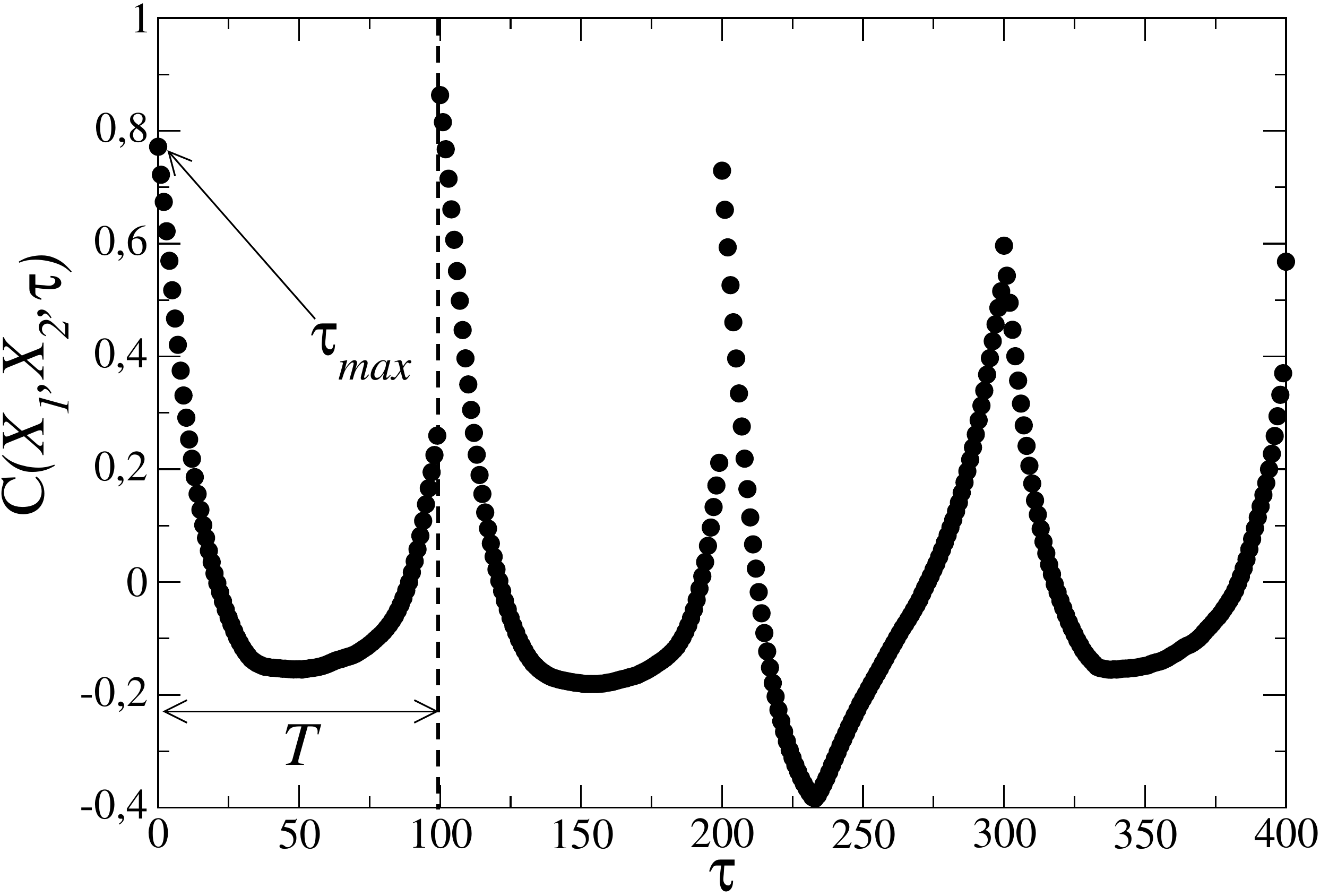}
\caption{\protect
         The 
         covariance in Eq.~(\ref{eq:cov}) between the two cells $1$ and
         $2$, shown in Fig.~\ref{fig3}, as a function of time-delay $\tau$.
         Here, $T$ is the time-window between successive stimuli.
         Typically, during each $T$ period, the covariance decreases
         from a maximum value to a minimum.
         The value of $\tau_{\max}$ giving the delay for which the 
         covariance is maximized is the lowest time-lag corresponding to 
         a relative maximum.}
\label{fig4}
\end{figure}

Figure \ref{fig4} shows the covariance between two astrocytes in the 
sample showed in Fig.~\ref{fig1}. 
Depending on the value of the delay $\tau$, the covariance between two
series can be large (close to $1$ or $-1$) or small (close to $0$).
The successive maxima shown in Fig.~\ref{fig4}
correspond to distinct stimuli.

However, the covariance alone does not suffice for showing 
how strongly connected two astrocytes are.
Indeed, the covariance for different 
delays correspond to different 
interaction modes, rates and paths joining two cells.
Here, we focus in the ``first'' interaction mode, corresponding
to the fastest path propagating the signal from the unique source-cell to 
each of the other cells in the tissue. We assume that detecting for
instance a reversible or irreversible effect of a drug in the
rapidity of these propagating paths is sufficient to determine
the reversible or irreversible effect in the tissue.

To determine the fastest paths, the time-delay in Eq.(\ref{eq:cov}) 
must also be considered, specifically the smallest one that maximizes the 
covariance: the delay needed for characterizing the connection between 
astrocytes is the lowest delay for which a local maximum of the 
covariance is observed. This value  will be referred to as $\tau_{max}$, 
cf.~Fig.~\ref{fig4},  and the corresponding covariance is  henceforth 
called $C_{max}=C(X_i,X_j,\tau_{max})$.

The time-delay $\tau_{max}$ in fact measures the typical time
for the signal to propagate from astrocyte $i$ to astrocyte $j$.
Notice that we do not consider the relative distances of the cells
because, as mentioned in the introduction above, the medium
in which the signals propagates is highly heterogeneous and 
therefore preferential and non-preferential pathways of signal 
propagation exist and the geometrical distance is not 
a suitable criterion for signal propagation velocity\cite{giaume1998}.

In the particular situation that a cell propagates its signal to two
neighbors with neither delay nor dumping, the respective covariances 
are exactly one, see Fig.~\ref{fig3}c. 
In this case both situations drawn with blue and
green arrows are equivalent and indistinguishable.
Either one, and only one, of them is known to connect the set 
of cells.

If the delay $\tau_{max}$ that maximizes the covariance between two 
astrocytes is small, it means that they should be 
closely connected and the corresponding connection should be strong.
If the time-delay is large compared to other pairs, the corresponding 
strength should be small compared to those other pairs.
Therefore, the weight of the connection is reasonably assumed to be
proportional to the inverse of the delay $\tau_{max}$.
 
Combining all the considerations above we define the strength $w_{ij}$ 
of the connection between astrocytes $i$ and $j$ as
\begin{eqnarray}
w_{ij}=\frac{\vert C_{max}(X_{i}, X_{j}, \tau_{max})\vert}{\tau_{max}+1} ,
\label{eq:W}
\end{eqnarray}
where one unit in the denominator is added for convenience,
to avoid singular behavior.
Notice that the weight $w_{ij}$, being the quotient
between correlation of two signals and time, 
can be interpreted as a correlation flux,
which in this case measures the causality -- and not the strength --
of the flow of information between cells.

Still, the weight value alone cannot reveal the structure of {\it single}
signals through the tissue, as we explain 
next\footnote{%
The full implemented algorithm can be shared for research purposes.
For that, please contact the authors.}.
\begin{figure}[t]
\centering
\includegraphics[width=0.5\textwidth]{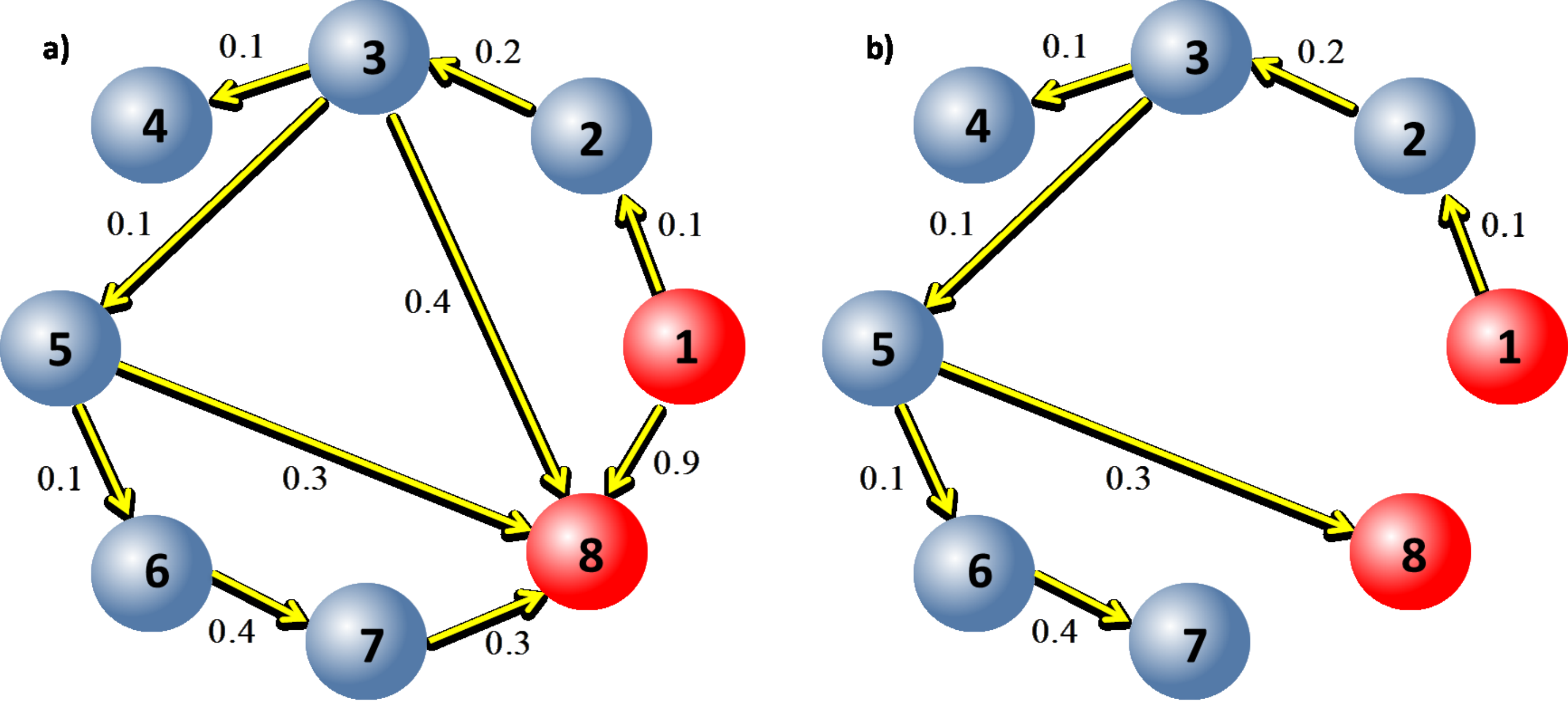}
\caption{\protect 
   Illustration of the computation of optimum paths.
   {\bf (a)}
   At each edge the corresponding delay $\tau_{max}$ is indicated.
   In this example the optimum path between cells $1$ and $8$ is 
   $P(1,8)=\{1,2,3,5,8\}$, because this path combines the least 
   time cost ($0.7$) and greatest number of edges ($4$). 
   See Tab.~\ref{tab1}.
   {\bf (b)}
   Consequently, the connections $\{1,8\}$, $\{3,8\}$ 
   and $\{7,8\}$ are redundant and therefore are filtered out by
   the procedure.}
\label{fig5}
\end{figure}
\begin{table}[htb]
\centering
\begin{tabular}{ccc}
\\
\hline 
Path & Time Cost & Number of Edges \\ 
\hline
 $\{1,8\}$ & $0.9$ & $1$ \\ 
 $\{1,2,3,8\}$ & $0.7$ & $4$ \\ 
 $\{1,2,3,5,8\}$ & $0.7$ & $5$ \\ 
 $\{1,2,3,5,6,7,8\}$ & $1.2$ & $7$ \\ 
\hline   
\end{tabular}
\caption{Time cost and number of edges of all paths between cells $1$ 
         and $8$ in the example sketched in \ref{fig5}.}
\label{tab1}
\end{table}
\begin{figure*}[htb]
\centering
\includegraphics[width=0.9\textwidth]{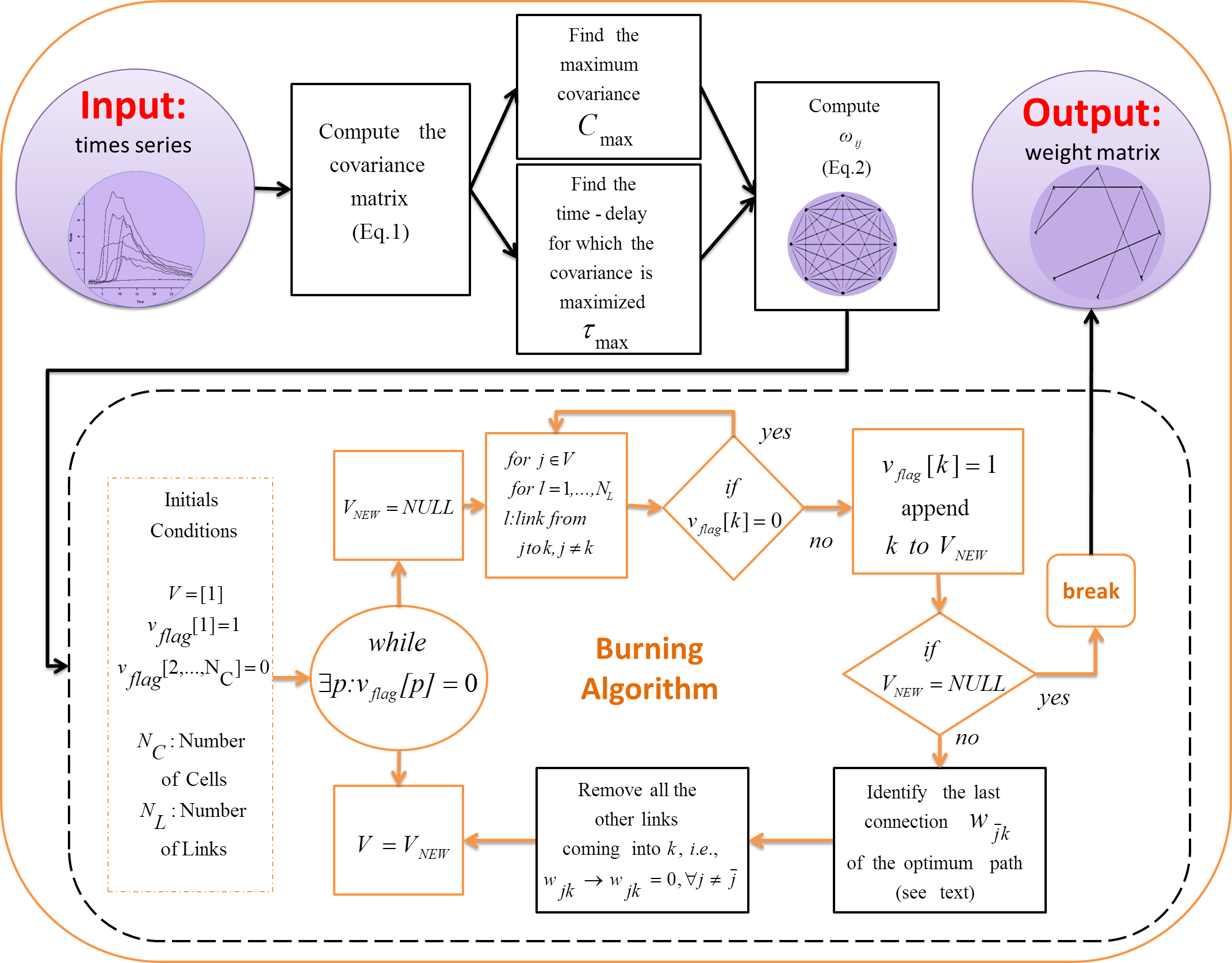}
\caption{\protect
         Infography 
         of the algorithm for extracting the connectivity
         network of astrocytic tissues (see Sec.\ref{sec:model}).}
\label{fig6}
\end{figure*}

\subsection{Constraints for signal propagation}
\label{subsec:contraints}

For any pair $i$ and $j$, the strength $w_{ij}$ is typically a non-zero 
value.
Therefore, by solely considering the values of $w_{ij}$ computed as in 
Eq.~(\ref{eq:W}), one cannot immediately infer the structure shown in
Fig.~\ref{fig3}a, since typically there is a non-zero covariance for
{\it any} pair of cells.
A final step is still necessary to filter out redundant connections.

The filter is based on two simple constraints for the connections.
First, there is one single source-cell characterized by having one or
more outgoing connections but no incoming connection. 
In all experiments shown, this is the first cell, $i=1$.
Second, all cells, different from the source-cell, have only one 
incoming connection, but can have several outgoing connections.
These two constraints are the sufficient and necessary ones for
cell-to-cell signal propagation, which is the situation we are
considering here.

From these two constraints the task reduces to extract the 
incoming functional connection of each cell that establishes an 
optimal (fastest) path from the source-cell to it.
We say that a path $P_O(i,j)$ between two cells, $i$ and $j$, is 
the optimum path between these cells if it has the minimum (total)
time-delay cost from all possibles path $P(i,j)$. In case one has 
more than one path corresponding to the minimum time-delay, one 
should choose the path maximizing the number of connections, in order
to minimize the time between each two adjacent cells in the path.
Figure \ref{fig5}, together with Tab.~\ref{tab1}, illustrates how the 
optimum path between each pair of cells is computed.
 
Starting from the source-cell, we traverse the network using a 
burning breadth-first algorithm\cite{sedgewick}:
start at a root node and inspect all its neighbors;
for each of those neighbors, inspect their neighbor 
nodes which were still not visited; and so on.
Then, for the source-cell we compute the optimum path from it 
to each one of its neighbors, removing all redundant incoming
connections of each neighboring cell. 
We iteratively repeat this procedure for each one of these neighbors, 
and therefore  for all cells. 
Figure \ref{fig6} summarizes the full algorithm.


\subsection{Verification of the network reconstruction algorithm with a simple model of \ca signal propagation }
\label{subsec:verification}

In order to numerically verify the reliability of our reconstruction algorithm, described in the previous sections
\ref{subsec:sig_corr} and \ref{subsec:contraints}, we will consider in this section
a synthetic network of cells, as sketched in Fig.~\ref{fig3}a, joined 
by directed connections (yellow arrows).  
We simulate the information flow through these artificial networks
by a simple auxiliary model of information flow described in detail in ~\ref{appendix:synth}.
At cell $1$ a Gaussian stimulus $X_1(t)$ is introduced.
Here, $X_1$ corresponds to the experimental ratio $R$ of both radiation amplitudes 
(see Fig.~\ref{fig1}) measured at cell $1$. 

By prescribing a time-delay to each connection, which
controls the necessary delay for the signal to propagate to the neighboring
cells, we extract the series of values composing the signal at each 
of the other cells.  Using the algorithm described in the previous subsections, 
we were able  to accurately uncover the connectivity  structure sketched in 
Fig.~\ref{fig3}a with yellow arrows solely by
analyzing the separated signals. In Fig.~\ref{fig3}b we indicate 
the result of our reconstruction with red arrows and compare it with results 
obtained from the standard  Granger algorithm for multivariate data\cite{blino2004} 
(black arrows).  Comparison with Fig.~\ref{fig3}a shows the accuracy of our reconstruction,
whereas the Granger model identifies spurious connections not
present in the network.
Next we carefully compared our algorithm with the standard Granger causality,
analyzing a set of $100$ artificial networks of $11$ synthetic cells ($10$ 
connections).
For details on the generation of synthetic data we again refer the reader 
to \ref{appendix:synth}, and for details on Granger causality procedures 
see \ref{appendix:granger}.

Figure \ref{fig7} shows the frequency of artificial networks that
correctly detect a given percentage of connections (efficiency).
As can be clearly seen, in all cases at least $50\%$ of the connections
were properly extracted, and typically the percentage of connections
correctly identified lies between $80\%$ and $90\%$. In comparison with
the standard Granger algorithm (red histogram), one concludes that for 
these signal-propagating networks the above procedure shows a high efficiency.
\begin{figure}[htb]
\centering
\includegraphics[width=0.45\textwidth]{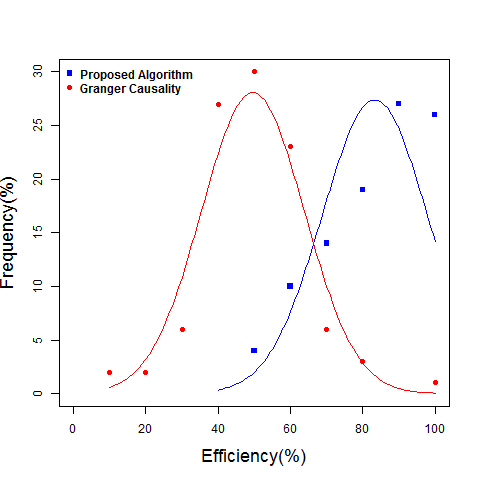}
\caption{\protect
         Testing the algorithm for extracting the connectivity network
         in living tissues of astrocytes. 
         For the procedure described in Sec.~\ref{sec:model}
         (blue histogram) all networks show a correctness larger than
         $50\%$ and most of them reach $80-90\%$. Such correctness is 
         significantly larger than the one obtained with standard Granger 
         causality (red histogram).
         The plot shows the percentage of connections correctly extracted 
         from $100$ artificial networks, each with $11$ cells and $10$ 
         connections.}
\label{fig7}
\end{figure}

Here efficiency $e$ was computed directly from the number 
$N_c$ of connections that were correctly predicted and the number of
connections $N_e$ equivalent to the original networks, as sketched
in Fig.~\ref{fig3}c, yielding $e=(N_c+N_e)/N_T$, being $N_T$ the total
number of connections correctly predicted. For our algorithm
the total number of connections equals the number of cells minus one,
$N_T=n_c-1$.
For the Granger algorithm $N_T$ is variable and lies between $n_e-1$ and
$n_e(n_e-1)$, since it is insensitive to the constraints for cell-to-cell
propagation introduced above.

It should be noted that there are two fundamental differences between 
Granger's method and ours.
First, our procedure considers a connectivity matrix weighted
by the time-lag between signals. Second, it introduces two constraints
necessary for uncovering the so-called primary graph in the particular
case of externally stimulated tissues of interconnected cells.
From these tissues one extracts multivariate signals that result 
from one single source signal -- the external stimulus -- 
which propagates throughout a spatially extended system.
By uncovering the primary graph, our procedure will not guarantee that
reversal connections do not exist. Still, focusing on the primary
graph, we see that 
it retrieves first order effects of the stimulus in the interconnected 
functional structure among cells propagating the signal, particularly in 
the case when a drug is used. 
We show below that these effects are complementary to the usual effects
uncovered through standard drug tests (see Fig.~\ref{fig8} below).
With our procedure, non-distinguishable connections are mutually exclusive, 
avoiding redundant connections. 

\section{Experimental setup and data extraction}
\label{sec:data}

As mentioned above in Sec.~\ref{sec:introducao}, activation of specific 
membrane receptors localized in the astrocytic plasma membrane triggers a 
rapid and brief rise of intracellular Ca$^{2+}$ concentration in this
cell, which promotes the release of gliotransmitters that will lead 
to the increase of intracellular Ca$^{2+}$ concentration on a 
neighboring astrocyte. 
Thus, stimulation of a single astrocyte in culture, with ATP 
(adenosine-5'-triphosphate), induces intracellular Ca$^{2+}$ elevation 
in the stimulated cell, which is then followed by Ca$^{2+}$ increases 
in neighboring astrocytes. 

The transmission of intercellular Ca$^{2+}$ signals between astrocytes is 
achieved through two distinct pathways: 
(i) release of gliotransmitters that 
will bind receptors located on neighboring astrocytes, and/or 
(ii) Ca$^{2+}$ itself, or a Ca$^{2+}$ liberating second messenger (as IP$_3$)
permeate gap junction channels and then act on similar 
intracellular targets in neighboring coupled cells.

\begin{figure*}[t]
\centering
\includegraphics[width=0.45\textwidth]{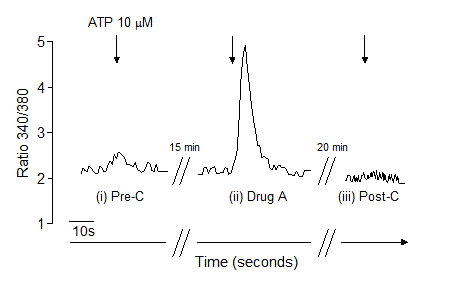}%
\includegraphics[width=0.45\textwidth]{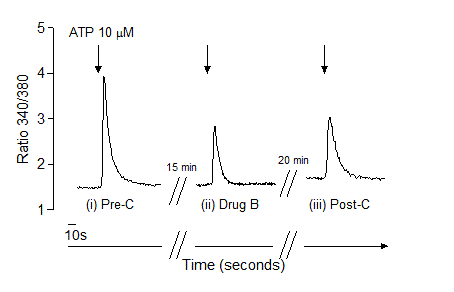}
\includegraphics[width=0.45\textwidth]{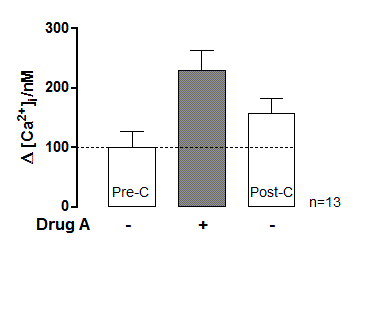}%
\includegraphics[width=0.45\textwidth]{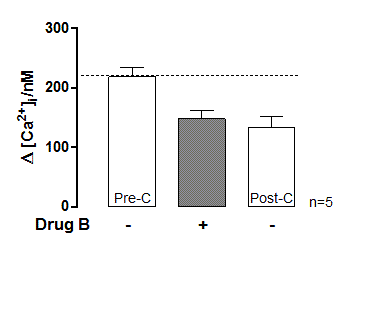}
\caption{\protect
         For determining if a drug is inhibitory or excitatory
         or to determine if its effects are reversible or not,
         one needs to observe the $R$ ratio for three different
         moments: (i) before applying the drug (Pre-C), 
         (ii) during the effect
         of the drug and after (iii) washing out the drug (Post-C).
         In the plots above
         we plot the $R$ amplitude for these three instants using
         two different drugs drug A (left) and drug B (right).
         While drug A shows an excitatory and reversible effect,
         drug B is inhibitory and irreversible. In the plots below
         the averages of Ca$^{2+}$ concentration for a total of ten 
         experiments are shown.
         Statistical significance of the drug effect
         is $P<0.05$. 
         Drug A:
              4-[2-[[6-Amino-9-(N-ethyl-$\beta$-D- 
              ribofuranuronamidosyl)-9H-purin-2-yl]amino]ethyl]benzene]propanoic acid hydrochloride;
              Drug B: N6-Cyclopentyl-9-$\beta$-D-Ribofuranosyl-9H-purin 
                      -6-amine.
         Throughout the experiments, ATP ($10\ \mu$M) was used as 
         the stimulus to evoque Ca$^{2+}$ signals. ATP was pressure
         applied for $200$ ms (arrows in (a) and (b)) through a micropipette
         placed over the cells.}
\label{fig8}
\end{figure*}

The calcium signal, as the ones recorded in the top plots of 
Fig.~\ref{fig8},
corresponds to a calcium signal by one cell, i.e.~the variation of the
fluorescence ratio, proportional to calcium concentration within the cell 
versus time. In a monolayer culture of astrocytes, such as the ones 
studied here,
the calcium signal propagates from one astrocyte to another creating a 
propagating calcium wave.
The propagation velocity of Ca$^{2+}$ waves reaches
$28.2$ $\mu$m/s\cite{newman2001} 
and it is also known that intracellular velocity ranges from $9.4$ to $61.2$ 
$\mu$m/s\cite{cornell1990}.
Thus, in order 
to have intracellular Ca$^{2+}$ waves\cite{chaospaper} that could be 
used to uncover the connections between the astrocytes we performed 
calcium imaging using primary cultures of cortical astrocytes.
For the results shown in Fig.~\ref{fig8} we have a statistical significance 
of $P<0.05$ (Student´s t-test) for the hypothesis that the (Gaussian)
distribution is the same
when the signal amplitude during drug perfusion, upon ATP stimulation,
is compared with the responses immediately before drug
perfusion (control).
\begin{figure*}[t]
\centering
\includegraphics[width=0.95\textwidth]{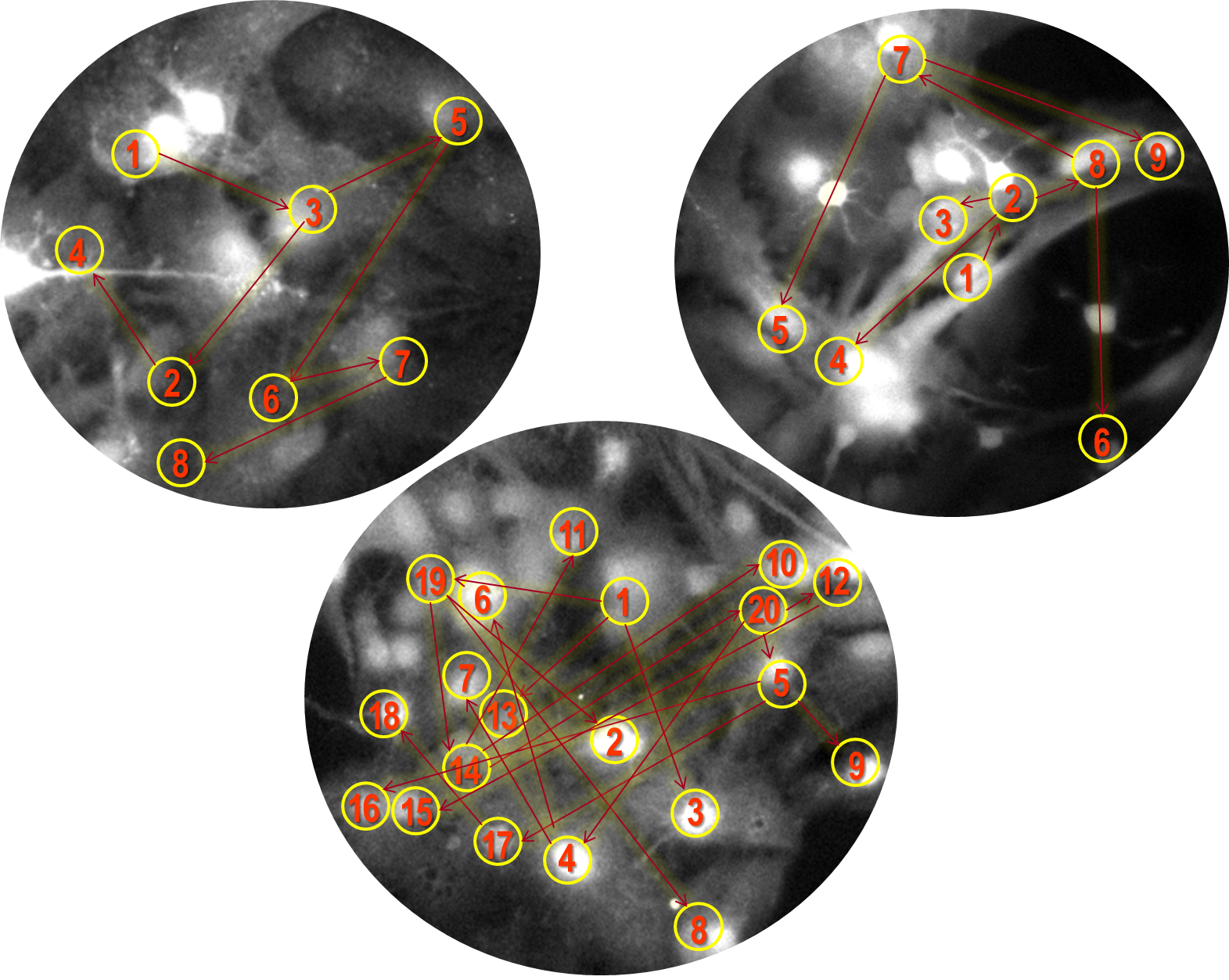}
\caption{\protect
         Three different astrocytic cultures
         with the corresponding derived network structure of Ca$^{2+}$
         signal propagation.
         Circles denote the location of spots where the signal
         was measured, namely in the cytoplasm.
         From the connectivity between cells and its evolution 
         through a succession of ATP stimuli, one is able to
         evaluate relevant  effects of the perfused drug on the 
         culture (see text).}
\label{fig9}
\end{figure*}

Primary cultures of cortical astrocytes from Wistar rats (0-2 days 
old) were prepared as reported previously\cite{vaz2011}
and in accordance with Portuguese laws and the European Union Directive 
86/609/EEC on the protection of Animals used for Experimental and other 
scientific purposes.

For calcium measurements, microglia contamination was minimized by 
following a standard shaking procedure\cite{mccarthy1980}. 
After six days in culture (DIC 6), cells in the T-75 culture flasks were 
shaken for $4$-$5$h at $37^o$C, the supernatant was removed and DMEM 
supplemented medium was added. At DIC 7, the T-75 culture flasks were 
shaken again for $2$-$3$ hours at $37^o$C, the supernatant containing 
mostly microglia was removed and then cells were washed once with PBS.
After removing microglia contamination, astrocytes to be used in calcium 
imaging experiments were plated (7x104 cells/ml) in $\gamma$-irradiated 
glass bottom microwell dishes.
Before plating, cells were gently detached by trypsinization 
(1\% trypsin-EDTA) for 2 minutes, the process being stopped by the addition 
of 4.5 g/l glucose DMEM medium containing 10\% fetal bovine serum with 
0.01\% antibiotic/antimycotic.

Astrocytes were loaded with the Ca$^{2+}$-sensitive fluorescent dye fura-2 
acetoxymethyl ester (fura-2 AM; 5µM) at 22$^o$C for 45 minutes. 
After loading, the cells were washed three times in external physiological 
solution (composition in mM: NaCl 125, KCl 3, NaH$_2$PO$_4$ 1.25, 
CaCl$_2$ 2, MgSO$_4$ 2, D(+)-glucose 10 and HEPES 10; pH 7.4 adjusted with 
NaOH)\cite{rose2003}.

Dishes were mounted on an inverted microscope with epifluorescent optics 
(Axiovert 135TV, Zeiss) equipped with a xenon lamp and band-pass filters of 
$340$ and $380$ nm wavelengths. Throughout the experiments, the cells were 
continuously superfused at $1.5$ ml/min with physiological solution
with the aid of a peristaltic pump. 
Calcium signals were induced by ATP,
applied focally, for $200$ ms, through a FemtoJet microinjector 
(Eppendorf, Hamburg, Germany), 
coupled to an ATP  ($10\ \mu$M) filled micropipette placed 
under visual guidance over a single astroglial cell.

In this study we address the two following drugs:
4-[2-[[6-Amino-9-(N-ethyl-$\beta$-D-%
ribofuranuronamidosyl)-9H-purin-2-yl]amino]ethyl]benzene]propanoic 
acid hydrochloride, henceforth named as ``Drug A'',
and 
N6-Cyclopentyl-9-$\beta$-D-Ribofuranosyl-9H-purin-6-amine,
henceforth named as ``Drug B''.
Both drugs were added to the external solution under perfusion. 
Changeover of solutions was performed by changing the inlet tube of the 
peristaltic pump from one flask to another; changeover of solutions with 
equal composition did not lead to appreciable changes of the responses. 
In each experiment and for each cell, responses to the stimulus (pressure 
applied ATP) were first obtained in the absence of the drug (control, Pre-C), 
then in the presence of the drug, after changeover of solutions, and lastly 
after returning to the drug free conditions (washout, Post-C).
Image pairs obtained every 250 ms by exciting the preparations at 340 
and 380 nm were taken to obtain ratio images. Excitation wavelengths 
were changed through a high speed wavelength switcher, Lambda DG-4 
(Sutter Instrument, Novato, CA, USA), and the emission wavelength was 
set to 510 nm. Image data were recorded with a cooled CCD camera 
(Photometrics CoolSNAP fx) and processed and analyzed using the software 
MetaFluor (Universal Imaging, West Chester, PA, USA). 

Regions of interest 
were defined manually over the cell profile. 
Typically one chooses the cytoplasmic region for measuring the 
ratio $R$, given preference to the brightest regions in each 
astrocyte appearing in the photo images during one stimulus.

It is important to know whether the effect of a certain drug
is inhibitory or excitatory, as well as whether it is reversible
or irreversible. To that end we compute the ratio $R$ (induced by
ATP application)
at three different time steps: before introducing the drug (left), in the 
presence of the drug (middle), and after washing out the effect 
of the drug (right). At each one of these moments one measures the magnitude
of $R$, having values respectively $R_{H_1}$, $R_D$ and $R_{H_2}$. 
If $R_D<R_{H_1}$ the drug 
has an inhibitory effect, while in the opposite case, $R_D>R_{H_1}$
the drug proves to be excitatory. In our case one sees that the effect
of drug A is excitatory (Fig.~\ref{fig8}, top left) while the effect of drug B
is inhibitory (Fig.~\ref{fig8}, top right).

To ascertain the reversibility of drug effects one takes 
in addition the magnitude $R_{H_2}$. If $R_D-R_{H_1}\sim R_D-R_{H_2}$
the effects are reversible. If not they should not be reversible,
yielding typically $R_D\sim R_{H_2}$. In the case illustrated in 
Fig.~\ref{fig8}, drug A is reversible while drug B is irreversible.

Notice that, the responses in left plot of Fig.~\ref{fig8} are from one cell.
Responses from the right plot are from another cell and another culture. 
The stimulation parameters in the condition illustrated in the top-left
plot of Fig.~\ref{fig8} 
were empirically adjusted (by changing the relative position of the 
stimulating electrode) to induce a weak response under control conditions 
since the protocol was designed to test the influence of a drug known to 
have excitatory actions.
Drug A is a well known and selective agonist of excitatory adenosine A2A 
receptors, known to be present in astrocytes\cite{vaz2011}. 
The stimulation parameters in the condition illustrated in top-right 
plot of Fig.~\ref{fig8} 
were empirically adjusted (by changing the relative position of the 
stimulating electrode) to induce a stronger response under control conditions 
since the protocol was designed to test the influence of a drug known to have 
inhibitory actions.
Drug B is a well known and selective agonist of inhibitory adenosine A1 
receptors, also known to be present in astrocytes\cite{cristovao2013}. 
Importantly, within each panel, the responses shown are all from the same 
cell, under exactly the same stimulation conditions.
Once set they were not changed up to the end of the recording period.
The difference being absence or presence of the drug in the perfusion 
solution.

As we will see in the next section, by assessing the 
connectivity network for each set of cells in one culture
of astrocytes we will be able to provide additional insight
to the effect of one drug in the living tissue.

\section{Assessing drug effects from cellular connectivity}
\label{sec:analysis}

The standard procedure described in Sec.~\ref{sec:data} for evaluating 
the excitatory and inhibitory effects of one drug or their reversible 
or irreversible character will in this section be extended to a broader 
context. Indeed, the approach done in the previous section considered the 
signal's total amplitude observed in the entire tissue sample.
Now, using the method introduced in Sec.~\ref{sec:model} allows us to 
retrieve the full structure of the connectivity network through which 
the injected stimulus propagates. 
In Fig.~\ref{fig9} three tissue samples are shown with their respective 
connectivity network for one particular stimulus.
Next, we apply this  procedure to a succession of ten stimuli
in each tissue sample shown in Fig.~\ref{fig9}.
For each stimulus, the connectivity network is defined by the weight 
matrix $w_{ij}$, quantifying the signal propagation between all 
sender cells $i$ and all receptor cells $j$.
\begin{figure*}[htb]
\centering
\includegraphics[width=0.24\textwidth]{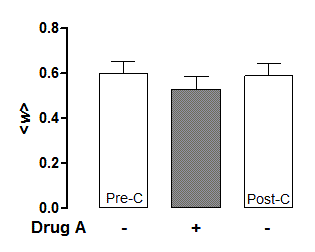}%
\includegraphics[width=0.24\textwidth]{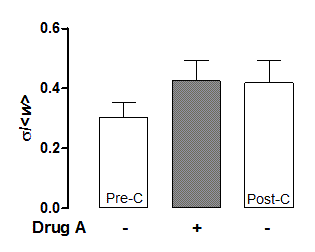}%
\includegraphics[width=0.24\textwidth]{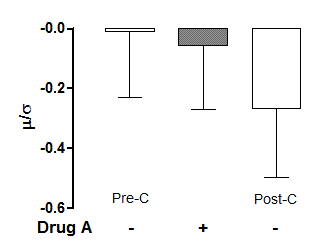}%
\includegraphics[width=0.24\textwidth]{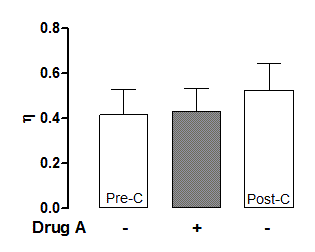}
\includegraphics[width=0.24\textwidth]{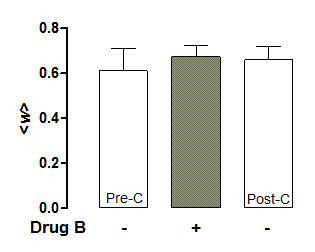}%
\includegraphics[width=0.24\textwidth]{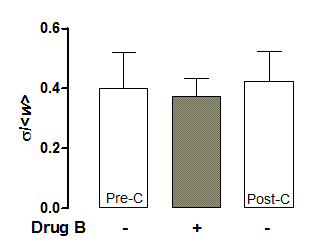}%
\includegraphics[width=0.24\textwidth]{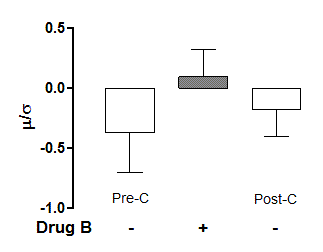}%
\includegraphics[width=0.24\textwidth]{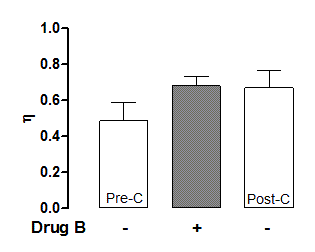}
\caption{\protect 
         Assessing the robustness and activity of 
         samples of astrocytic tissues throughout a series of ten 
         stimuli: two initial stimuli (Pre-C), four drug A (first row)
         or drug B (second row) and four 
         washout (Post-C).
         Drug A: 
              4-[2-[[6-Amino-9-(N-ethyl-$\beta$-D- 
              ribofuranuronamidosyl)-9H-purin-2-yl]amino]ethyl]benzene]propanoic acid hydrochloride;
              Drug B: N6-Cyclopentyl-9-$\beta$-D-Ribofuranosyl-9H-purin 
                      -6-amine. 
         From left to right one sees result for:
         flux $\langle w\rangle$ (first column) in Eq.~(\ref{mean}),
         sensitivity $\sigma/\langle w\rangle$ (second column) in Eq.~(\ref{sigma}),
         strengthening $\mu /\sigma$ (third column) in Eq.~(\ref{mu}),
         variability $\eta$ (fourth column) in Eq.~(\ref{persistence}).
         Here we use a total of ten experiments.}
\label{fig10}
\end{figure*}

For instance, 
while a drug may have an inhibitory effect on the amplitude, reducing 
the overall signal strength, it may simultaneously change the signal 
propagation
network in a way that the signal, though weaker, propagates more easily,
i.e.~it has a facilitatory effect in the propagating structure.
We introduce four additional quantities, each one of them reflecting the
facilitatory or inhibitory -- and reversible or irreversible 
-- effects induced by a particular drug. 
As does $w_{ij}$, these additional properties
characterize not the intensity of the signal but the stronger or weaker
ability of the signal to propagate throughout the tissue, i.e.~the
causality of the information flow. Therefore,
they can be taken as properties complementary to the amplitude.

The four additional quantities are all computed directly from the
weight $w_{ij}$ introduced above, using the auxiliary quantity 
$\bar{w}_{ij}=(1/10)\sum_{T=1}^{10}w_{ij}(T)$ ($T=1,\dots,10$), the 
average of the weight between each pair of cells, over the ten 
experimental phases.
By ascertaining how the weights $w_{ij}$ change from
one stimulus to the next, we are able to determine the effects of 
the substance in the tissue, which is reflected by the moments of 
the weight distribution.

The first moment of the distribution is simply 
\begin{equation}
\langle w\rangle(T) = \frac{1}{L}\sum_{i\neq j} {w}_{ij}(T),
\label{mean}
\end{equation}
with $L$ indicating the total number of connections.
While the set of values $\bar{w}_{ij}$ represents the time-average
strength of one single connection in time,
$\langle w\rangle(T)$ indicates the average weight -- or flux --
per connection in the tissue for a particular stimulus at time T.

The second moment is important for ascertaining how influential
is a particular drug in inducing a variation of the weight
between two connected cells.
It is computed by accounting for the fluctuations 
around the means $\bar{w}_{ij}$ and averaging them
over the $L$ connections:
\begin{equation}
\sigma^2 (T) = \frac{1}{L-1}\sum_{i\neq j} ({w}_{ij}(T)-\bar{w}_{ij})^2.
\label{sigma}
\end{equation}
When $\sigma=0$ it implies that the connectivity network is precisely
the same for all stimuli and therefore the drug has no influence
in the weights. 
The larger the value of $\sigma$ is, the stronger the influence of the
drug to induce a variation in each connection.
Typically, large fluctuations are more probable when the average 
weight is also large. Therefore, to remove this scaling effect, 
we consider the
normalized second moment, $\sigma/\langle w\rangle$, which quantifies 
the fluctuations with respect to the observed average weight.
We call this coefficient the sensitivity coefficient.

For evaluating how powerful a drug is in weakening or strengthening
the connectivity between each pair of cells we consider the third
moment of the weight distribution:
\begin{equation}
\mu^3 (T) = \frac{1}{L}\sum_{i\neq j} ({w}_{ij}(T)-\bar{w}_{ij})^3.
\label{mu}
\end{equation}
When $\mu=0$ it means that the amount of connections with a strength
below average is the same as the amount of connections with a strength
above it. If $\mu\neq 0$ the weight distribution is asymmetric.
A positive value indicates that the values of the weights concentrate on 
the right-side of the distribution, i.e.~there are few weak connections,
while a negative value indicates that there are few strong connections.
Consequently, when a stimulus leads to a connectivity network with a
larger (smaller) $\mu$ value than previously, it has a strengthening
(weakening) effect on the connectivity of the tissue. 
Similarly to the second moment,
we consider $\mu/\sigma$, normalized to
the standard deviation $\sigma$. We call this coefficient the
strengthening coefficient.

A fourth measure is added to these three moments, which we
call variability $\eta$. It is a function of the 
stimulus $T$ that evaluates how much the weights between each pair of
cells varies from one stimulus to the next one:
\begin{equation}
\eta(T)=\frac{\sum_{i,j}\vert w_{i,j}(T+1)-w_{i,j}(T)\vert}{\sum_{i,j}\vert w_{i,j}(T+1)+w_{i,j}(T)\vert} .
\label{persistence}
\end{equation} 
The variability takes values between zero, when all connections
remain the same from one stimulus to the next one, and one,
when all connections switch from zero at $T$ to one at $T+1$ or
vice-versa. The larger the variability the broader the 
overall induced change in the connectivity network.

For each tissue sample we considered a succession of typically ten
stimuli, similarly to what was done above. 
The first two stimuli were applied in the absence of drugs (buffer) 
in order to uncover the tissue connectivity when not subjected to drugs. 
Then one of two drugs, herein referred to as drug A and drug B, were applied 
and the astrocytes stimulated four times.
Drugs were then removed from the bath and the astrocytes stimulated again 
four times (washout period).

\begin{table}[t]
\centering
\begin{tabular}{ccc}
\\
\hline 
Property & Drug A & Drug B \\ 
\hline
Amplitude & $+$ (Rev.) & $-$ (Irrev.) \\
(Fig.~\ref{fig8}) & & \\
\hline
Flux $\langle w\rangle$ & $0$ & $0$ \\
(Fig.~\ref{fig10}, 1st col) & & \\
\hline
Sensitivity $\sigma/\langle w\rangle$ & $+$ (Irrev.) & 0 \\
(Fig.~\ref{fig10}, 2nd col) & & \\
\hline
Strengthening $\mu/\sigma$ & $0$ & $0$ \\
(Fig.~\ref{fig10}, 3rd col) & & \\
\hline
Variability $\eta$ & $0$ & $+$ (Irrev.) \\
(Fig.~\ref{fig10}, 4th col) & & \\
\hline   
Correlation $\rho$ & $+$ (Irrev.) & $+$ (Rev.) \\
(Fig.~\ref{fig11}) & & \\
\hline
\end{tabular}
\caption{\protect
         Table of evaluation properties for the drugs, drug A and
         drug B, in signal amplitude and the properties assessing
         the structure of the signal propagation network. For
         each property we indicate whether its effects are 
         facilitatory ($+$) or inhibitory ($-$) and 
         reversible (Rev.) or irreversible (Irrev.).
         Zero indicates no statistical significance in a T-test
         among a total of ten experiments.
         Last row shows the correlation (see text), which deviates
         from the purely random case $\rho=1/4$, except when applying 
         drug B (see Fig.~\ref{fig11}).}
\label{tab2}
\end{table}
\begin{figure}[htb]
\centering
\includegraphics[width=0.24\textwidth]{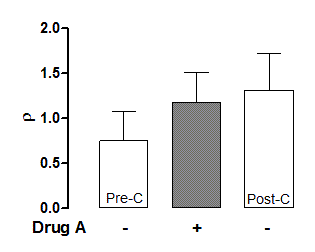}%
\includegraphics[width=0.24\textwidth]{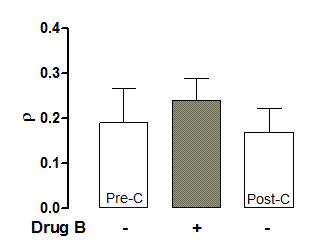}
\caption{\protect 
  Correlations for drug A (top) and drug B (bottom)
  as shown in Eq.~(\ref{correlation}). For the period
  when applying drug B one observes $\rho\sim 1/4$
  meaning that the functional structure is random.
  Still, before and after the application of drug B,
  the correlation deviate from the purely random case.
  The same occurs for the experiments with drug A
  (see text).}
\label{fig11}
\end{figure}

Figure \ref{fig10} shows the four coefficients, 
$\langle w\rangle$, $\sigma/\langle w\rangle$, $\mu/\sigma$ 
and $\eta$,
measured for each stimulus.
As for measuring the R ratio in Fig.8, we define a drug effect as 
inhibitory if $R_D < R_{H_{1}}$, or as exhibitory if $R_D > R_{H_{1}}$, 
and the classification as reversible or irreversible is based on 
the comparison of $R_D- R_{H_{1}}$ and $R_D - R_{H_{2}}$.
In Tab.~\ref{tab2} we present a summary of the main results 
from Fig.~\ref{fig10}, where ``$0$'' indicates no statistical
significance in a T-test ($p>0.1$).
As one can see, both the flux and the strengthening are not
affected by any of the drugs, while the variability is.
However, differently from the amplitude, variability
shows to be affected by drug B, being 
facilitatory and irreversible. 
Drug A also shows a irreversible
and facilitatory effect on the variability but curiously 
with some  delay. This may indicate that the time for
drug application, while being properly used when evaluating
its effect in total signal amplitude (Fig.~\ref{fig8})
may be not long enough for other features of the tissue
influencing the signal propagation, in this case the
variability.
Such increase of the variability could be a sign that the
structure supporting some robustness of signal propagation
is permanently modified by each drug. 
The same facilitatory and irreversible effect on the sensitivity 
seems to also occur, but only for drug A.
It deserves to be noted that in a completely random
model of link creation in each time step $T$, the variability would
be peaked at a value of $1/2$.
This is not supported by our data, which is
a strong indicator that the reconstructed network do not
have overwhelmingly random character.

For these new properties, we may summarize the
analysis in this section by stating that while
drug A mainly affects sensitivity $\sigma/\langle w\rangle$,
drug B mainly affects variability $\eta$.

Comparing 
the variability $\eta$ with the amplitude one 
concludes that, at least for these two 
different properties and for the correlation measure,
the effects of the drugs are of different nature. Therefore,
one should approach such drug effects in a more extended way
to analyze not only its overall influence in one single 
property, such as the total amplitude of the signals, 
but also in a set of properties that combined, characterize
the functional structure of the network.

Finally, it is important to check wether the functional
structure does not change randomly among the several experiments
under the same conditions.
Figure \ref{fig11} shows the one-step correlation
\begin{equation}
\rho(T) = \frac{1}{L} \sum_{i\neq j} w_{ij}(T+1)w_{ij}(T) 
\label{correlation}
\end{equation}
for all three periods when testing drug A (left) and drug B (right).
If we imagine a simple binomial model in which functional connections 
between cells $i,j$  are randomly created in each step, i.e. 
$w_{ij}(t) = ran(0,1)$, both correlations would be equal to $1/4$. 
Correlation values above $1/4$ indicate the presence of actual 
correlations and memory about previous correlations in the system.
When applying drug B correlations approximate the uncorrelated 
regime $\rho=1/4$. For both pre and post periods however, correlation
deviates significantly from $\rho=1/4$, evidencing a functional 
structure underlying the sequence of signals. For drug A this
deviation is even stronger and it is further strengthened when
applying the drug in a irreversible way (check Tab.~\ref{tab1}).

\section{Conclusions}
\label{sec:conclusions}

In this paper we introduce a procedure for extracting the 
functional connectivity network in living tissues subjected to external pulsed 
stimuli  that propagate through it as a signal.
Our method is based on the covariance matrix of the separated
signals taken at different time-lags. By adding proper constraints
of a minimum time-delay between pairs of cells and single-source 
stimulus to each cell we are able to filter out all redundant 
or artificial connections
from the covariance matrix.
We test our procedure with synthetic data. There, our procedure 
proves to be better suited for assessing the 
connectivity of living tissues of   astrocytic samples used
than other standard measures, namely the 
multivariate Granger causality algorithm.
The better results of our procedure indicates that for the
particular case of signal propagation networks with one
single triggering source, standard methods may retrieve
biased results.

The weight (strength) of each connection is computed directly
from the covariance matrix and the minimum delay-time. 
Further, we showed how to obtain additional insight relative to 
the drug used for stimulating the source-cell only by analyzing 
the distribution of connectivity strengths (weights).
From the first, second and third moment of the weight distribution
we showed how to characterize respectively the signal flux and also 
the sensitivity and strengthening of the network underlying the 
signal propagation in the tissue.
We also characterized the temporal stability of the network by its 
variability and a correlation measure, and found these measures to 
complement the information from the signal amplitude ratios.

Following what was introduced in the beginning of our paper,
it is important to stress here that our method is not able to
uncover the physical connectivity structure between cells and 
that it is also not aimed for studying particular features of the 
signal itself, such as the measurement noise\cite{perc2008}.
While our method works for pulsed signals similar to the ones measured
in the samples shown in Fig.~\ref{figAppend}, as explained in 
\ref{appendix:synth}, and the variability
already reflects implicitly the impact of measurement noise in the sample
of signals, it would be interesting to extend
this methodology further to situations where the measurement noise
is significant.

To further validate our method, it would be interesting to test it
with an empirical known drug whose effect in the propagating structure 
is known. To our knowledge, there is no such drug test, since
they typically focus on the overall (sum) signal in the tissue and not 
in the features of its propagation throughout the tissue.

All in all, this study proposes a procedure complementary to the
standard approach where only the overall amplitude is tested 
before, during and after the application of a given drug. Although
it is not able to extract the physical interconnections between 
cells, our simple procedure provides a way for quantifying 
the functional connectivity in the tissue and to ascertain how 
it changes due to the application of drugs.
The framework  introduced here as well as the reported
findings should now be used for a systematic pharmacological study.
Other more sophisticated approaches, namely a spectral 
analysis\cite{spectral} applied to these signal-propagation networks, 
could be useful. In this case larger tissues, having a larger number 
of astrocytes, are needed. 

Finally, while the method introduced in this paper is based in constraints
that give results for primary functional connections, some other interactions 
with a real physiological meaning could be further considered.
Namely, there can be autocatalytic phenomena by self-stimulation, some 
signals could be reversible or multiple stimulation of one cell may occur
in other situations. 
Forthcoming studies could further improve the ability for characterizing 
drug effects in living tissues by considering these more general effects, 
i.e.~several time-lags, or local maxima of the matrix solution, so 
that a hierarchy of interactions could be determined.

\section*{Acknowledgments}

The authors thank Matthew Blow for his suggestions on the text.
MP thanks {\it Ci\^encia sem Fronteiras}, Brasil, 
Ref.~MCT/CNPq/CsF/202611/2012-4.
Financial support from {\it Funda\c{c}\~ao para a Ci\^encia e a Tecnologia}, 
Portugal is acknowledged by FR (SFRH/BPD/65427/2009),
SHV (SFRH/BPD/81627/2011), PGL ({\it Ci\^encia 2007}) and also partial 
support under PEst-OE/FIS/UI0618/2011 and the binational program between 
FCT and DAAD, DRI/DAAD/1208/2013 by PGL and FR.


\appendix
\section{Generation of synthetic data}
\label{appendix:synth}

In order to evaluate our  method for reconstruction the connectivity network, 
we propose a simple signal propagation model with properties similar to 
those found in the experiment. 

The model for signal propagation treats a set of nodes $i$, $i=1,\dots,M$ 
with respective simulated signal time series $\hat{X}_i(t)$ that are linked 
through connection with strength (weight) $w_{ij}$, either $0$ or $1$. 

The signal propagation is governed by the following rules:

\begin{itemize}

\item[i)] The first node, $i=1$, is driven by an external Gaussian signal 
          $\hat{X}_1(t) \propto \exp  \left( -\frac{(t-t_0)^2}{2 \sigma_0^2} \right)$,
          where $t_0$ is the starting time and $\sigma_0$ is the standard deviation.
          Here we choose $t_0=30$ and $\sigma=3$.

 \item[ii)] Each node $i\neq 1$ has only one incoming connection $j$ ($j \neq i$), 
            i.~e.~only one nonvanishing $w_{ji}$.

 \item[iii)] There are no simple loops, i.~e.~$w_{ii}=0$, for all $i$.

 \item[iv) ] At each time step $t$ the signal departing from cell $i$ is 
             distributed
             through its outgoing connections transporting the signal to its
             neighbors $j$. The distribution is implicitly defined as:
             \begin{equation}
                 \hat{X}_i(t) =\alpha \sum_j w_{ij} \hat{X}_j(t+\tau_{ij}) \,,
             \end{equation}
             with $\alpha=1/(1+\sum_j w_{ij})$ and the delay $\tau_{ij} = 1$ if 
             $w_{ij}=1$ or $\tau_{ij} = 0$ otherwise.
 \end{itemize}

\begin{figure}[t]
\centering
\includegraphics[width=0.43\textwidth]{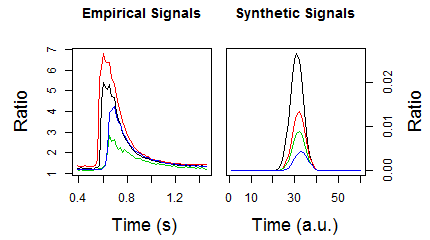}
\caption{\protect 
         Comparison between the empirical signals measured on
         astrocytic tissues and synthetic signals generated as
         described in Sec.~\ref{appendix:synth}.}
\label{figAppend}
\end{figure}

This means that the cells receive the incoming signal which is delayed by 
an amount in this case given by the strength of the incoming connections 
themselves.
Such synthetic signals, while not having exactly the same shape as the
empirical signals, do reproduce attenuation and stimulus delay 
in a similar way (see Fig.~\ref{figAppend}).

\section{Granger causality}
\label{appendix:granger}

Granger causality procedures aim to test whether one time series is 
able to forecast another one. 
The main idea can easily be explained by considering, for simplicity,
two time series $X_1$ and $X_2$. 
One says that $X_1$ Granger-cause $X_2$ if the series of values of
$X_1$ provide information about future values of $X_2$.
More precisely, we select a proper autoregression of $X_2$ which has
say $m$ previous values
\begin{equation}
X^{(1)}_2(t)=a_0+\sum_{i=1}^m a_iX_2(t-i)+\epsilon^{(1)}_t,
\end{equation}
and one computes the estimate errors $\epsilon_t$.

This autoregression is then compared with another one where the 
values of $X_1$ are considered:
\begin{equation}
X^{(2)}_2(t)=a_0+\sum_{i=1}^m a_iX_2(t-i)+\sum_{i=1}^m b_iX_1(t-i)+
\epsilon^{(2)}_t.
\end{equation}
If the latter estimate error $\epsilon^{(2)}_t$ is smaller then the 
former one $\epsilon^{(1)}_t$, $X_1$ is assumed to Granger-cause $X_2$.

In our paper we considered an improved multivariate version 
of Granger causality, where the $M$ time-series are taken 
simultaneously
and the error estimates of autoregressions are now for all pairs of 
variables and then compared as a matrix. Following the procedure 
outlined in \cite{blino2004}, we fit to our vector of measured 
time series $\mathbf{X}(t) = [X_1(t), X_2(t), \dots, X_M(t)]^T$ a MVAR (multivariate auto-regressive) model as
\begin{equation}
  \label{eq:MVAR}
  \mathbf{X}(t) = \sum_{k=1}^{p} \mathbf{A}(k) \mathbf{X}(t-k) + \mathbf{E}(t)
\end{equation}
with a delay dependent coefficient matrix $\mathbf{A}(k)$, a given maximum delay $p$, and an error matrix $\mathbf{E}$, which are assumed to be Gaussian white noise sources. Transforming the quantities to the domain of frequencies $f$, one can compute the so-called transfer matrix $\mathbf{H}(f)$ from the relation
\begin{equation}
  \label{eq:transfer_matrix}
  \mathbf{A}^{-1}(f) \mathbf{E}(f) = \mathbf{H}(f) \mathbf{E}(f) \,,
\end{equation}
from which one can compute the directed transfer function (DTF) $\gamma$ as 
\begin{equation}
  \label{eq:dtf}
   \gamma_{ij}^2 (f) = \frac{|H_{ij}(f)|^2}{\sum_{m=1}^M |H_{im}(f)|^2} \,.
\end{equation}

In a network of nodes $i,j$,  the DTF is a measure of signal causality for the propagation of the signal from node $j$ to node $i$. In our case, we use a single delay $p=1$, effectively eliminating the frequency dependency of the DTF, which reduces to a matrix of connection strengths. It should be noted, however, that the multivariate Granger approach assumes stationarity of the underlying time series in order for the MVAR mechanism to be applicable.  Stationarity does not hold in our case, as the measured signals are non-stochastic soliton-like peaks with a compact support.
This is the fundamental reason for Granger causality to fail when applied to stimuli signals in living tissues.



\end{document}